\documentclass[conference]{IEEEtran}

\usepackage[switch]{lineno}

\IEEEoverridecommandlockouts
\usepackage[most]{tcolorbox}
\definecolor{darkorange}{RGB}{255, 140, 0}
\definecolor{darkblue}{RGB}{84, 112, 198}

\usepackage{cite}
\usepackage{amsmath,amssymb,amsfonts}
\usepackage{algorithmic}
\usepackage{graphicx}
\usepackage{textcomp}
\usepackage{xcolor}
\def\BibTeX{{\rm B\kern-.05em{\sc i\kern-.025em b}\kern-.08em
    T\kern-.1667em\lower.7ex\hbox{E}\kern-.125emX}}

\usepackage{booktabs}
\usepackage{multirow}
\usepackage{array}
\usepackage{graphicx}
\usepackage{subcaption}

\usepackage{listings}
\usepackage{caption}
\usepackage{xspace}
\PassOptionsToPackage{most}{tcolorbox}
\newcommand{\systemname}{Proteus\xspace}
\DeclareRobustCommand{\IEEEauthorrefmark}[1]{\smash{\textsuperscript{\footnotesize #1}}}
\begin{document}

\title{Rethinking Burst Buffer Optimization: Enabling Layout Heterogeneity via Hybrid Analysis and LLM Guidance\\

}

\author{
    \IEEEauthorblockN{
        Yuhan Cai\IEEEauthorrefmark{1},
        Huijun Wu\IEEEauthorrefmark{1},
        Zhuo Tang\IEEEauthorrefmark{2},
        Kehua Guo\IEEEauthorrefmark{3},
        Wenzhe Zhang\IEEEauthorrefmark{1},
        Zhenwei Wu\IEEEauthorrefmark{1},\\
        Zhouyang Jia\IEEEauthorrefmark{1},
        Ruibo Wang\IEEEauthorrefmark{1}, and
        Yong Dong\IEEEauthorrefmark{1}
    }
    \IEEEauthorblockA{
        \IEEEauthorrefmark{1}National University of Defense Technology, Changsha, Hunan, China
    }
    \IEEEauthorblockA{
        \IEEEauthorrefmark{2}Hunan University, Changsha, Hunan, China
    }
    \IEEEauthorblockA{
        \IEEEauthorrefmark{3}Central South University, Changsha, Hunan, China
    }
    \IEEEauthorblockA{
        Email: \{caiyuhan, wuhuijun\}@nudt.edu.cn
    }
}

\maketitle

\begin{abstract}
Burst buffers (BBs) are essential for mitigating I/O bottlenecks in modern HPC systems. However, existing BB file systems often suffer from structural performance degradation due to fixed data layouts that fail to align with diverse application behaviors. While current machine-learning-based optimizations focus primarily on tuning storage stack parameters for a given layout, they offer diminishing returns when a fundamental mismatch exists between I/O patterns and the underlying data organization. Furthermore, these approaches typically incur prohibitive costs due to extensive training or intrusive profiling.

To bridge this gap, we present \systemname{}, a semantic-aware BB system that treats data layout as a first-class optimization dimension. The core insight of \systemname{} is that application I/O intent can be reconstructed by synergetically combining static code structures with lightweight runtime signals. Through a hybrid pipeline and a single execution probe, \systemname{} extracts latent semantic cues to determine the optimal layout prior to production runs—eliminating the need for prior training or exhaustive profiling. Evaluation with representative HPC workloads shows that \systemname{} achieves 91.30\% decision accuracy, delivering up to 3.24$\times$ and 2.9$\times$ speedups for write-intensive and metadata-intensive workloads, respectively.
\end{abstract}


\section{Introduction}

As high-performance computing (HPC) systems scale toward exascale, the I/O subsystem has increasingly become the dominant factor limiting end-to-end application efficiency~\cite{liu2012role, oral2019end}. While compute capabilities continue to grow rapidly, storage systems advance at a much slower pace in terms of bandwidth, latency, and scalability, forcing applications to idle during I/O phases~\cite{xie2012characterizing}. Burst buffers (BBs)~\cite{liu2012role, wang2014burstmem} have thus been widely adopted as intermediate storage layers to absorb transient I/O bursts and decouple applications from backend parallel file systems (PFS). 

Despite their effectiveness, most existing BB systems (e.g., GekkoFS~\cite{gekkofs}, UnifyFS~\cite{unifyfs}, CodepFS~\cite{CodepFS}) embed a critical yet often overlooked assumption: application I/O behaviors are semantically uniform. In practice, a BB must commit to a specific data and metadata layout, access path, and routing strategy at job startup, and this choice remains fixed throughout execution. As a result, system performance is tightly bound to a single assumed I/O semantic.

This rigidity becomes a fundamental limitation under heterogeneous and multi-phase HPC workloads. When an application’s actual I/O intent diverges from the assumed layout, the resulting performance degradation is structural rather than incidental. Such inefficiencies arise from inherent mismatches between application semantics and storage architecture, and therefore cannot be mitigated through parameter tuning or conventional runtime optimizations. These observations suggest that supporting multiple layout modes and selecting the appropriate one for each workload is essential for unlocking the full potential of burst buffer systems.

The central challenge lies in accurately and efficiently identifying application I/O behavior prior to execution. Existing machine-learning (ML) based approaches face significant obstacles in production HPC environments. They rely on large volumes of historical I/O traces and continuous retraining, incurring substantial storage, computational, and operational overheads\cite{MLinsuper}. In addition, these approaches typically require repeated application executions to collect sufficient training data, while many production workloads run only a limited number of times, making such data collection impractical\cite{mlhpc22,IOmlsurvey}. Furthermore, they operate within a fixed architecture and are confined to tuning within a fixed BB data layout, leaving fundamental structural layout mismatches unresolved.

To enable more practical and generalizable burst buffer optimization, we adopt a different perspective: performance depends on understanding the application’s I/O intent before execution. Such intent is primarily encoded in static artifacts (e.g., source code structure, I/O API usage, and job script configuration), and can be complemented by lightweight runtime signals to resolve behavioral ambiguities.We therefore propose a hybrid approach that combines static analysis, lightweight dynamic signals, and large language models (LLMs) to guide burst buffer optimization with minimal data and execution overhead.

Based on this insight, we present \systemname{}, a semantic-aware burst buffer system that enables robust layout selection through hybrid intent inference. \systemname{} integrates three complementary components: (1) static intent extraction from source code and job scripts, (2) lightweight runtime profiling (e.g., a single execution probe via Darshan) to capture unambiguous dynamic behaviors, and (3) LLM-based semantic reasoning that jointly interprets these static and dynamic contexts in a zero-shot manner to generate the layout decision. Unlike prior work, \systemname{} avoids heavy runtime monitoring and does not require massive training data or repetitive parameter sweeps, while significantly improving decision accuracy.
Architecturally, \systemname{} extends GekkoFS to support four data and metadata layout modes that span the burst buffer design space. Layout decisions are applied at the job level, avoiding costly online reconfiguration and data migration.

This work makes the following contributions:

\begin{itemize}
    \item \textbf{Architectural Contribution.} We systematically identify and implement a multi-mode burst buffer system that unifies diverse data and metadata layouts within a single framework, enabling flexible adaptation to heterogeneous I/O semantics.
    
    \item \textbf{Methodological Contribution.} We propose a hybrid intent inference framework that combines static analysis and lightweight runtime evidence with LLM-based semantic reasoning, significantly improving robustness while avoiding the overhead of heavy runtime profiling.

    
    
    \item \textbf{System Evaluation.} Through extensive experiments on representative HPC workloads, we demonstrate that \systemname{} improves layout selection accuracy and achieves significant performance gains across diverse scenarios with minimal overhead.
\end{itemize}

\section{Background and Motivation }
\label{sec:background}

The performance challenge of burst buffers has shifted from whether acceleration is possible to under what conditions it can be achieved. In practice, performance varies widely across applications and I/O phases due to implicit assumptions about I/O semantics embedded in BB architectures. Data and metadata placement therefore define the performance envelope of a BB system, and any single static layout inevitably mismatches some workloads, creating performance ceilings that cannot be resolved through tuning. Multi-mode layouts are thus required to expand semantic coverage, raising two key questions: how to design such modes and how to select them before execution without runtime trial and error.

\subsection{Structural Limits of Burst Buffers Imposed by Data/Metadata Layout}
\label{sec:layout}

As computational performance scales, I/O bottlenecks increasingly constrain HPC workloads.
Early studies on leadership-class systems revealed severe checkpoint and output bottlenecks, where backend parallel file systems cannot sustain application concurrency and burstiness~\cite{xie2012characterizing}.
Recent studies further show that I/O is not a marginal overhead: many HPC applications spend 15$\sim$40\% of their execution time performing I/O~\cite{macedo2024behind}, and I/O interference under congestion can reduce application I/O throughput by up to 67\%~\cite{gainaru2015scheduling}.
The problem is even more pronounced in large-scale AI workflows, where data I/O can consume up to 90\% of total training time~\cite{pumma2019scalable}.

Burst buffers (BBs) have been widely deployed as an intermediate storage layer between compute nodes and parallel file systems~\cite{liu2012role,Tiered}, leveraging node-local or near-node storage devices such as SSDs and NVRAM to absorb bursty traffic, decouple application I/O from backend contention, and improve throughput~\cite{liu2012role,DaleyGLDRW20}.
Many BB designs have been proposed, often claiming performance advantages~\cite{wang2014burstmem,BBdataAPP15}.
However, their effectiveness is not determined by device bandwidth alone, but fundamentally depends on how data and metadata are placed, managed, and routed under different workload semantics.

\begin{figure}[t]
    \centering
    \includegraphics[width=0.85\columnwidth]{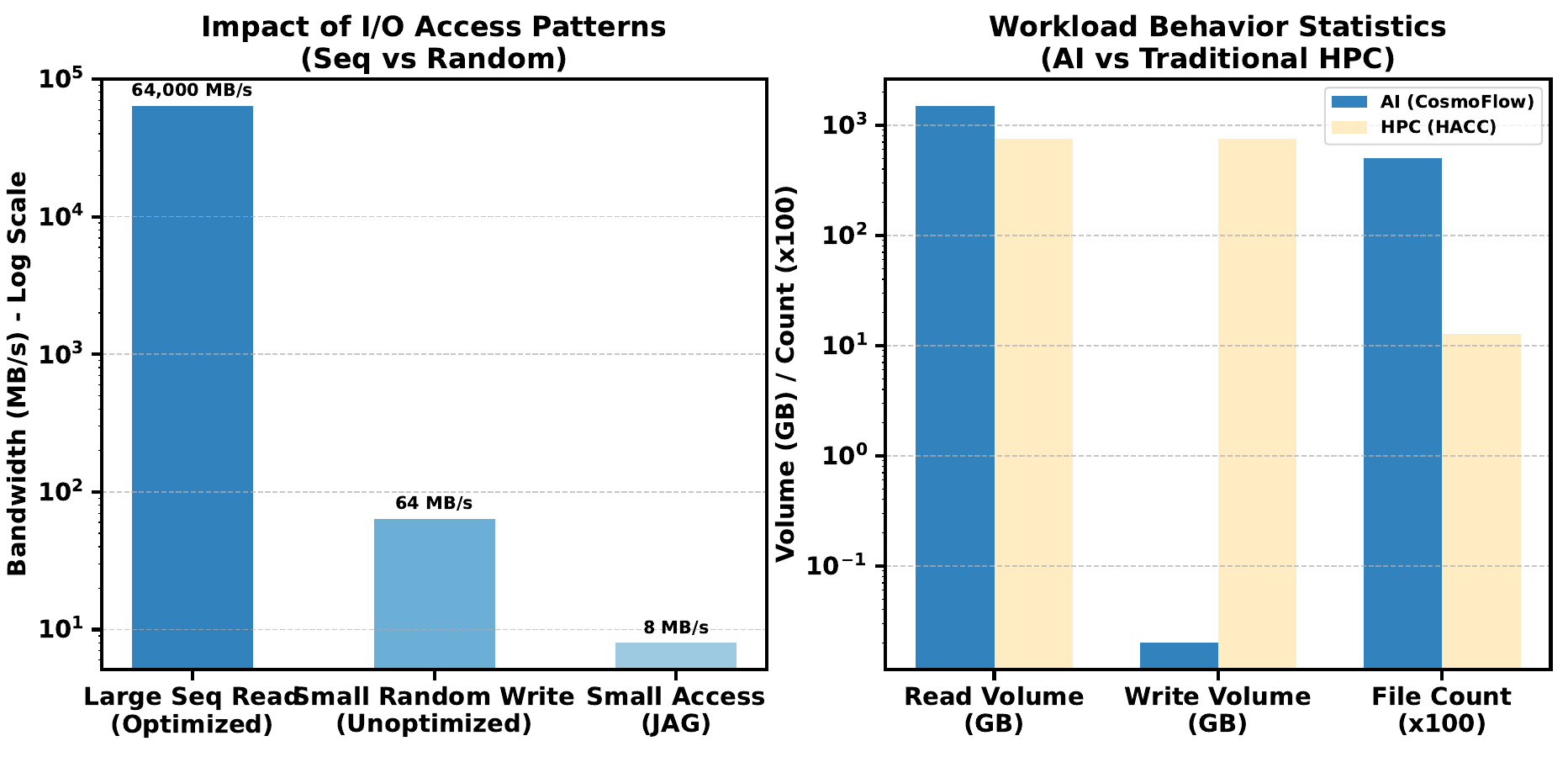}
    \caption{I/O heterogeneity across workloads~\cite{IObehavior22}. (a) The significant bandwidth gap between sequential and random access; (b) Distinct I/O characteristics of AI versus HPC workloads. These variations show that no single static layout fits all scenarios.}
    \label{fig:io_pattern_and_workload}
\end{figure}

Even on identical hardware, I/O patterns show dramatically different performance (Figure~\ref{fig:io_pattern_and_workload}).
A BB’s data and metadata layout encodes assumptions about application I/O semantics.
For example, GekkoFS uses a stateless distributed hashing scheme favoring concurrent, independent accesses but incurring overhead for global sequential scans~\cite{BBplacement19}.
In contrast, UnifyFS and HadaFS accelerate I/O through node-local writes, implicitly assuming write-dominant workloads with strong locality and limited cross-node access~\cite{hadafs,unifyfs}.
When workloads involve frequent reads, sharing, or collective access, these assumptions can degrade performance.
This demonstrates a fundamental reality: no single BB architecture is optimal across all I/O semantics, because its data and metadata layout determines which workloads it favors.

More specifically, layout encodes several critical trade-offs that shape these preferences.
First, \textbf{locality vs. sharing}: extreme locality maximizes bandwidth but isolates data, hindering cross-node coordination; global sharing improves coordination but adds latency due to distributed locks or routing.
Second, \textbf{metadata coupling vs. scalability}: co-locating metadata with data enables low-latency access but can trigger RPC storms under complex cross-directory patterns; distributed metadata improves scalability but incurs coordination overhead.
Finally, \textbf{access aggregation strategy} affects performance: direct paths minimize latency but may underutilize the network, whereas intermediate aggregation boosts throughput under high load at the cost of control complexity~\cite{BBdesign15,BBplacement19}.

\begin{figure}[t]
    \centering
    \includegraphics[width=0.98\columnwidth]{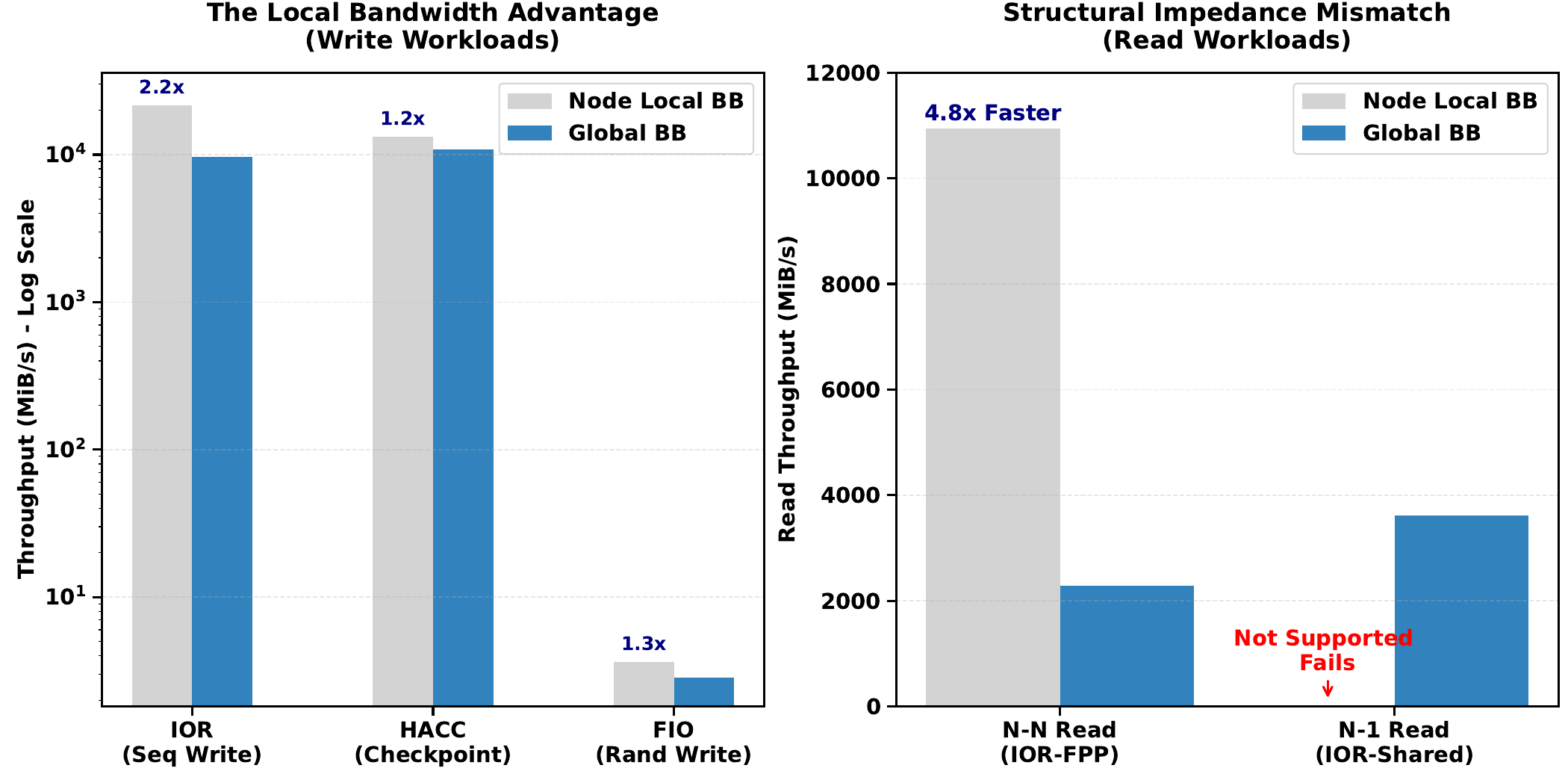}
    \caption{Performance impact of layout selection. Node-local layouts excel at write-intensive checkpoints, while global layouts are required for read-heavy or shared-file workloads, illustrating the penalty of layout mismatch.}
    \label{fig:local_vs_global_bb}
\end{figure}

These trade-offs define the structural performance envelope of BB systems.
Any fixed layout embeds assumptions about workload behavior, and deviations from these assumptions impose a structural performance ceiling.
Parameter tuning or runtime optimization can barely overcome these limitations, as shown in Figure~\ref{fig:local_vs_global_bb}.

To address this, \textbf{Proteus} adopts a multi-mode design with data layout as the primary axis of distinction.
It provides four modes differing in data and metadata placement and access paths, allowing the BB to reshape its architecture prior to job execution based on semantic workload features.
This approach establishes layout as central to performance optimization and lays the foundation for LLM-driven mode selection.

\subsection{Burst Buffer Layout Selection: Challenges and Opportunities}
\label{sec:preexecution}

As discussed above, the layout of a burst buffer (BB) directly determines the achievable performance ceiling, making it natural to adapt data and metadata placement based on application I/O patterns. In practice, however, realizing this strategy remains challenging.

A common approach is to leverage dynamic instrumentation tools, such as Darshan or Recorder, to capture runtime I/O behavior and build performance models that guide configuration selection~\cite{OPRAEL,predictionI/O23,IOaitunning24}. However, their applicability is limited in production HPC environments. Building reliable models requires substantial training data and repeated executions under varying configurations, incurring prohibitive profiling overhead and making them impractical for workloads with limited execution runs.

Although historically difficult to exploit, recent studies have demonstrated that static analysis can effectively characterize performance-relevant I/O properties without requiring full execution \cite{Static17,FQL19,CodeIO18}. This suggests that static information provides a lightweight and semantically rich foundation for understanding the logical structure of application I/O intent.
However, comprehensive I/O characterization requires understanding not only the logical structure (e.g., concurrency, access patterns) but also the execution intensity (e.g., exact read/write volumes, multi-phase durations). While static analysis provides accurate structural baseline, it struggles with edge cases involving dynamic loops or input-dependent I/O sizes (e.g., exact read/write volumes and multi-phase durations). Conversely, dynamic profiling captures these empirical metrics accurately but suffers from the severe overheads discussed earlier.
To achieve both completeness and efficiency, we propose a mutually complementary hybrid approach. We combine the structural insights extracted from static artifacts with empirical evidence collected from lightweight runtime probes. This design effectively captures both dimensions of I/O behavior to disambiguate uncertain cases, while introducing minimal optimization tax.

Recent advances~\cite{cwm,codeio} in large language models (LLMs) further enable effective integration of these heterogeneous signals. In \textbf{Proteus}, the LLM serves as a semantic reasoning engine that jointly interprets static artifacts and lightweight runtime evidence, mapping inferred I/O intent to structured layout modes. This hybrid design enables robust and efficient layout selection across diverse workloads, achieving improved decision quality without incurring significant runtime overhead.

\begin{figure*}[h]
    \centering
    \includegraphics[width=0.95\textwidth]{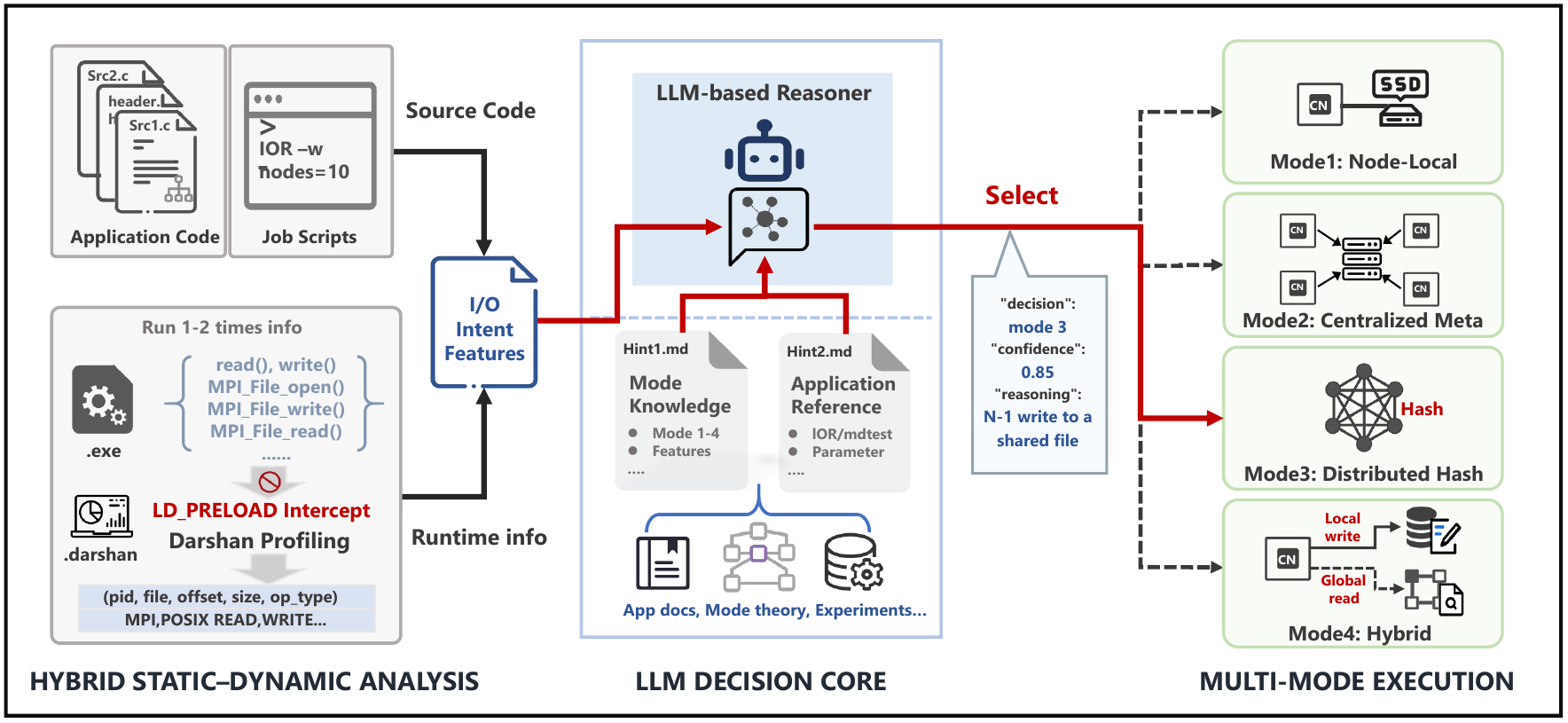}
    \caption{
The system performs hybrid I/O intent extraction from static artifacts and lightweight runtime signals, uses LLMs to select the optimal burst buffer layout, and instantiates the underlying storage architecture prior to execution.}
    \label{fig:system_architecture}
\end{figure*}

\section{System Design}
\subsection{System Overview}
\label{sec:system_overview}
\systemname is designed to address the structural mismatch between application I/O intent and rigid storage layouts.
As illustrated in Figure~\ref{fig:system_architecture}, the framework operates at the job submission stage, leveraging both static application artifacts and lightweight runtime signals to determine an appropriate burst buffer configuration.
The design goal is to achieve accurate and robust layout orchestration with minimal runtime overhead. The workflow consists of three primary phases:

\begin{enumerate}

\item \textbf{Hybrid intent extraction.}
This phase captures the application’s I/O behavior by combining two complementary sources: static artifacts (e.g., source code and job scripts) and lightweight runtime signals collected from a single execution probe. The outputs of both sources are consolidated into a unified structured profile (i.e., the hybrid context).

\item \textbf{LLM-based semantic reasoning.}
Operating on the extracted hybrid context, the decision core leverages an LLM guided by storage domain knowledge. The LLM semantically correlates the application's I/O characteristics with the architectural trade-offs of different layouts to select the optimal layout mode.

\item \textbf{Multi-mode layout activation.}
Based on the inferred decision, the system activates the selected layout mode prior to job execution. By instantiating routing rules and placement policies, the system ensures that the application operates within an optimized storage configuration from the outset, without requiring runtime reconfiguration.

\end{enumerate}


\subsection{Multi-Mode Burst Buffer Architecture}
\label{sec:multi_mode_engine}

In high-performance computing environments, achieving layout diversity through hardware customization or specialized interconnect topologies is both cost-prohibitive and inherently non-portable. 
The core design principle of \systemname{} is therefore to implement multiple storage layouts on top of a single physical architecture. 
By introducing a logic-defined and configurable routing layer, \systemname{} enables the system to adapt its storage architecture at job granularity, without requiring physical data migration or hardware reconfiguration.

To characterize the structural alignment between application I/O semantics and burst buffer layouts, we analyze representative systems, including Cray DataWarp, BeeGFS, GekkoFS, UnifyFS, and HadaFS, as concrete instantiations of distinct points in a shared design space. 
This analysis reveals three fundamental and often conflicting dimensions: (1) locality versus sharing, (2) data--metadata coupling, and (3) namespace consistency guarantees. 
Based on these observations, \systemname{} adopts a multi-mode layout design that systematically spans this space through four complementary modes, covering workloads from isolated $N$--$N$ checkpointing to shared $N$--$1$ access patterns.
Rather than restricting optimization to parameter tuning within a single layout, \systemname{} exposes multiple fundamentally different layout strategies, enabling dynamic layout selection based on the specific I/O semantics of each workload.

To realize this design, \systemname{} abstracts complex storage layouts into a logical composition of three core routing functions. 
I/O requests are intercepted at the client-side layer, where a mode-specific routing table determines the destination host $\mathcal{H}$ before entering the distributed storage stack:
\begin{itemize}
    \item \textbf{Data placement function:}
    \[
        f_{\textit{data}}(path, chunk\_id) \rightarrow \mathcal{H}_i,
    \]
    determining the storage node responsible for each data chunk.

    \item \textbf{File metadata function:}
    \[
        f_{\textit{meta\_f}}(path) \rightarrow \mathcal{H}_i,
    \]
    determining the owner node for file-level metadata.

    \item \textbf{Directory metadata function:}
    \[
        f_{\textit{meta\_d}}(path) \rightarrow \{\mathcal{H}_{id}, \dots\},
    \]
    determining the scope of directory-level metadata management.
\end{itemize}

Routing decisions are implemented using high-efficiency function pointers, maintaining constant $O(1)$ lookup complexity and negligible overhead compared to network and storage latencies. 
By specializing the routing function triplet $\langle f_{\textit{data}}, f_{\textit{meta\_f}}, f_{\textit{meta\_d}} \rangle$, \systemname{} realizes distinct architectural templates without modifying the storage backend or introducing separate execution engines. 
Each mode corresponds to a structured operating point in the burst buffer design space, obtained through explicit trade-offs among locality, sharing, data–metadata coupling, and consistency.

\paragraph{Mode 1 (Node-Local Storage)}
Mode~1 represents the extreme end of the locality spectrum, prioritizing node-local access over global visibility.
All routing functions are deterministically mapped to the local node:
\[
f_{\textit{data}}
=
f_{\textit{meta\_f}}
=
f_{\textit{meta\_d}
}
\;\rightarrow\;
\texttt{localhost}.
\]
At the logical level, this eliminates all remote routing and effectively bypasses the RPC protocol stack.
Operations such as \texttt{forward\_write} are resolved as local synchronous calls, preserving strictly local data and metadata ownership.
This design removes network contention and metadata coordination entirely, making it well-suited for large-scale independent $N$--$N$ write workloads, such as checkpoint phases in tightly coupled simulations.
Systems such as Cray DataWarp’s private mode~\cite{datawarp} exemplify this design point, though Mode~1 in \systemname{} is derived as a general architectural extreme rather than a system-specific optimization.

\paragraph{Mode 2 (Centralized Metadata).}
Mode~2 occupies the opposite end of the sharing dimension by enforcing a globally consistent namespace through centralized metadata management.
A configurable subset of nodes, denoted as $S_{md}$
is designated as metadata servers and controlled by the parameter \texttt{metadata\_server\_ratio}.
The file metadata routing function is redirected to this subset according to
\[
f_{\textit{meta\_f}}(path)
\;\rightarrow\;
\texttt{str\_hash}(path)
\bmod |S_{md}|,
\]
while data placement remains distributed across all storage nodes.
By decoupling data distribution from metadata ownership, Mode~2 provides a strongly consistent global view of the namespace, significantly reducing path resolution overheads in $N$--$1$ shared access patterns and metadata-intensive workloads.
This mode corresponds to a design point commonly observed in global file systems such as BeeGFS~\cite{beegfs}, but is here generalized as a controllable architectural choice within a unified routing framework.

\paragraph{Mode 3 (Distributed Hashing)}
Mode~3 targets scalability by fully decentralizing both data and metadata routing.
All routing functions are implemented using consistent hashing, yielding deterministic and coordination-free placement decisions.
During \texttt{forward\_write}, block-level hashing is applied as
\[
\hspace{-1mm} f_{\textit{data}}(path, \! chunk\_id) \! \rightarrow \! \texttt{str\_hash}(path \! \Vert \! chunk\_id) \! \bmod \! N. \hspace{-1mm}
\]
uniformly distributing data across the cluster.
This design minimizes metadata hotspots and enables near-linear scalability under high concurrency, at the cost of weaker namespace semantics and limited locality guarantees.
Mode~3 serves as the load-balancing baseline in \systemname{}, providing robust parallel throughput for unstructured or random I/O workloads.
Distributed file systems such as GekkoFS~\cite{gekkofs} illustrate this design point, which in \systemname{} is realized purely through routing specialization rather than a dedicated execution path.

\paragraph{Mode 4 (Hybrid).}
Mode~4 explores an asymmetric routing strategy that decouples write-time locality from read-time global accessibility.
During the write phase, the data routing function is forced to resolve to the local node via a cached mapping:
\[
f_{\textit{data}}(path)
\;\rightarrow\;
\texttt{pathhost\_}[path].
\]
This preserves node-local write bandwidth and avoids network contention.
In contrast, file metadata routing remains globally distributed using hashing:
\[
f_{\textit{meta\_f}}(path)
\;\rightarrow\;
\texttt{str\_hash}(path) \bmod N.
\]
Metadata entries record an explicit \texttt{data\_location\_rank} field, enabling transparent redirection during subsequent cross-node reads.
This hybrid design targets multi-phase workflows characterized by private data generation followed by collective analysis, capturing a common pattern in modern HPC and data-driven pipelines.
Systems such as HadaFS~\cite{hadafs} exemplify this asymmetric design point, which \systemname{} generalizes within its routing-based abstraction.

By realizing layout diversity through logic-defined routing, Proteus decouples layout policies from the underlying protocol implementation. Routing rules explicitly dictate synchronization (e.g., triggering global coordination in Mode 2, or bypassing it via node-local paths in Mode 1). This enables seamless switching across diverse storage architectures without modifying the backend mechanisms.


\subsection{Hybrid Intent Inference Pipeline}
\label{sec:static_intent_and_reasoning}

\begin{figure}[t]
    \centering
    \includegraphics[width=1.2\columnwidth]{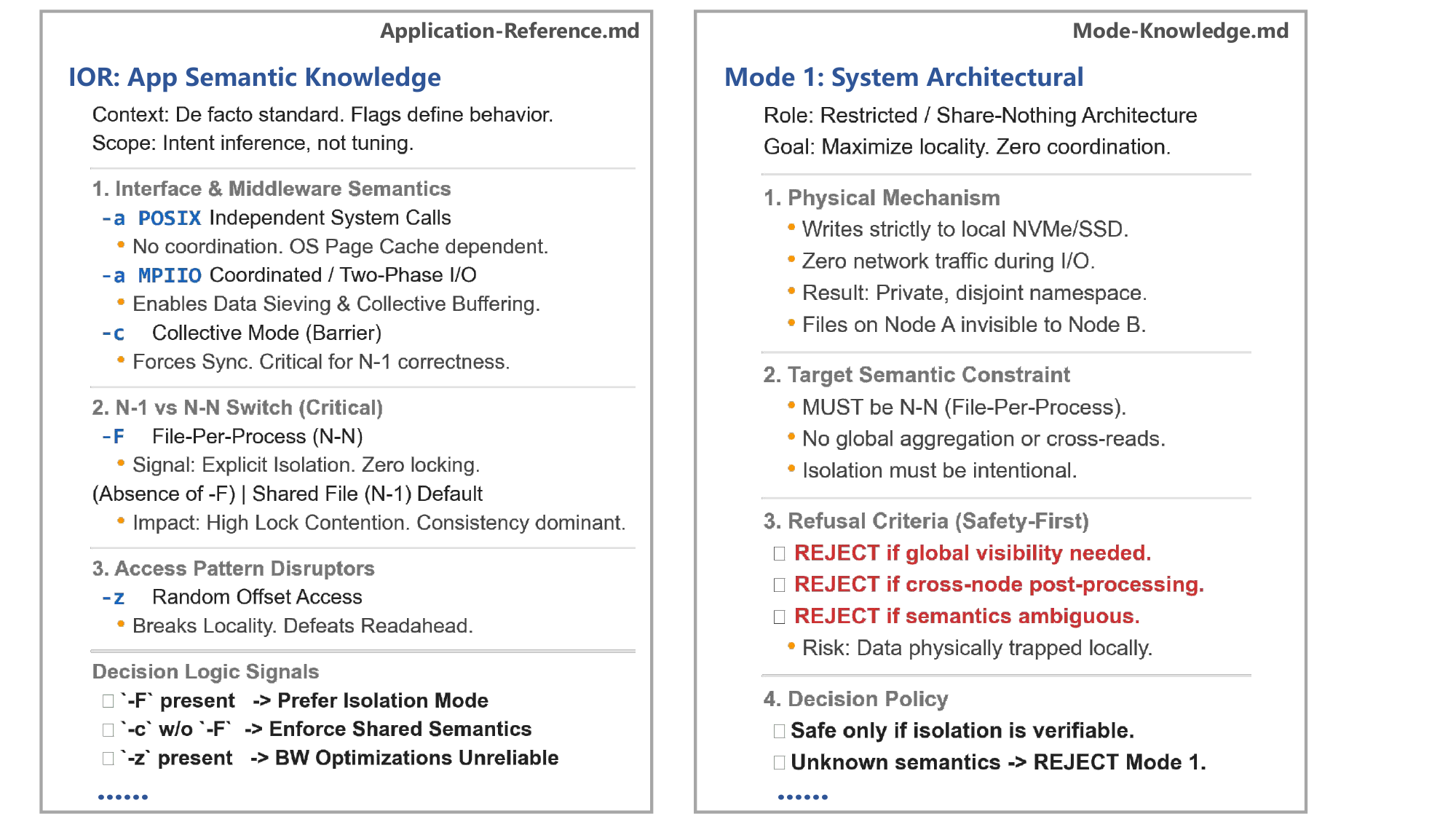}
     \caption{Examples of the domain knowledge base, encompassing application-level I/O semantics and mode-level architectural knowledge.}
    \label{fig:lmm_knowledge_base}
\end{figure}

While the multi-mode architecture defined in Section~\ref{sec:multi_mode_engine} establishes the structural potential for optimization, unlocking this potential requires a precise decision mechanism. As discussed in Section~\ref{sec:preexecution}, traditional ``optimize-by-running'' approaches suffer from high cost, poor generalization, and cold-start limitations in HPC environments. 

To accurately select the optimal layout, \systemname{} must understand the application's true I/O intent. Crucially, static and dynamic signals provide complementary dimensions of I/O behavior: static artifacts (e.g., source code and job scripts) reveal the logical I/O topology and concurrency structures, whereas lightweight runtime profiles supply empirical execution metrics (e.g., read/write ratios and metadata intensity). Therefore, \systemname{} captures both dimensions by combining compiler-based static extraction with lightweight runtime profiling, forming a comprehensive hybrid context. Operating on this unified representation, the LLM reasoning engine infers key I/O characteristics and maps them to an appropriate layout mode. Based on this design, layout selection is performed through a structured pipeline:

\paragraph{Context Extraction}

\systemname{} extracts a compact hybrid context to support burst-buffer layout selection. This stage focuses on gathering evidence directly relevant to data and metadata placement by combining static analysis with lightweight runtime profiling.

On the static side, the static extractor analyzes both source code and job scripts. From source code, \systemname{} identifies I/O call sites, file-name construction patterns, MPI rank usage, collective I/O operations, and rank-dependent control flow. These signals help distinguish common access topologies. For example, rank-indexed file names indicate a file-per-process N--N pattern, whereas multiple ranks accessing the same path with collective MPI-IO suggests an N--1 shared-file pattern. From job scripts, \systemname{} recovers launch parameters, process counts, benchmark options, file paths, transfer sizes, and sharing flags, because many HPC applications expose their I/O behavior through execution-time configuration.

On the dynamic side, \systemname{} performs a lightweight Darshan-based probe run to compensate for potential insufficiencies in the static context. Importantly, this probe is not used to search over candidate layouts or tune the system through repeated execution. Instead, it collects only behavioral summaries such as read/write ratio, dominant request size, metadata intensity, access regularity, and shared-file activity.

Finally, \systemname{} combines source-derived hints, script-derived parameters, and runtime summaries into a structured hybrid context that records evidence across layout-relevant dimensions, including access topology, file sharing, request granularity, metadata pressure and collective I/O usage (Figure~\ref{fig:hybrid_json}).

\begin{figure}[t]
    \centering
    \begin{tcolorbox}[
        colback=gray!5,
        colframe=black!75,
        title=\scriptsize\textbf{Hybrid Context},
        fonttitle=\bfseries,
        boxrule=0.6pt,
        arc=2pt,
        left=3pt, right=3pt, top=3pt, bottom=3pt,
        width=0.98\columnwidth
    ]
    \ttfamily\scriptsize

\textcolor{blue}{// Script-derived}\\
"bench\_params": \{\\
\quad "-a": "POSIX",\\
\quad "-b": "128m",\\
\quad "-t": "4m",\\
\quad "-o": "shared\_file"\\
\quad ... \\
\},\\[2pt]

\textcolor{magenta}{// Source-derived}\\
"static\_features": \{\\
\quad "access\_pattern": "strided",\\
\quad "topology\_hint": "N-1",\\
\quad "collective\_io": true\\
\quad ... \\
\},\\[2pt]

\textcolor{teal}{// Runtime (Darshan)}\\
"runtime\_stats": \{\\
\quad "posix\_bytes\_written": "5.2GB",\\
\quad "posix\_bytes\_read": "128MB",\\
\quad "posix\_meta\_ops": 24,\\
\quad "posix\_seq\_access\_ratio": 0.87\\
\quad ... \\
\},\\[2pt]


    \end{tcolorbox}
    \caption{Hybrid context combining script, source, and runtime (Darshan) features.}
    \label{fig:hybrid_json}
\end{figure}

\paragraph{Knowledge-Augmented LLM Reasoning}

To guide the LLM's reasoning, \systemname{} maintains a domain knowledge base covering two categories of information: (1) application-level semantics  and (2) mode-level descriptions, as illustrated in Figure~\ref{fig:lmm_knowledge_base}. The former captures I/O behaviors of common middleware and benchmarks, while the latter defines the architectural strengths and trade-offs of each \systemname{} layout.
By providing summarized application patterns and layout trade-offs, LLMs are equipped with decision-critical guidelines for accurate layout selection.

\paragraph{LLM-Driven Layout Selection}

Given the hybrid context and the domain knowledge, \systemname{} performs layout selection through structured reasoning over explicit I/O features. As illustrated in Figure~\ref{lst:prompt_design}, the model consumes a structured prompt that unifies the outputs of the previous steps: the extracted hybrid context and the domain knowledge. Unlike traditional machine learning models that merely fit decision boundaries from data without semantic constraints, making them highly dependent on data quality and scale, LLMs have internalized vast amounts of code, system documentation, and academic research~\cite{grattafiori2024llama,yang2025qwen3}. This inherent knowledge implicitly bridges the reasoning link from application I/O patterns to system design features and their matching relations. Essentially, using the LLM fundamentally eliminates the need for massive training data while introducing an explicit chain of reasoning. Specifically, the prompt explicitly enforces a step-by-step derivation path, requiring the LLM to sequentially analyze critical dimensions: concurrency topology (isolated vs. shared), resource intensity (metadata vs. bandwidth), I/O direction, and phase behavior. This structured reasoning prevents model hallucination and ensures that the decision is grounded in a transparent understanding of the workload’s I/O physics rather than empirical correlation. This structured output encompasses the selected mode, a confidence score, and a justification. Upon parsing, the system backend utilizes this decision to pre-provision the optimal storage layout, thereby enabling the multi-mode layout activation engine detailed in Section~\ref{sec:multi_mode_engine} to instantiate the appropriate configuration prior to job launch.

To ensure system stability and prevent catastrophic execution errors (e.g., stranded local data if Mode 1 is mistakenly applied to a shared-read phase), Proteus employs a fallback strategy. The system activates specialized architectural modes only when the LLM generates a high-confidence semantic match. In cases of behavioral ambiguity or low confidence scores, Proteus defaults to the robust Mode 3 (Distributed Hashing) as a fail-safe baseline.

\begin{figure}[t]
    \centering
    \begin{tcolorbox}[
        colback=gray!5,
        colframe=black!75,
        title=\scriptsize\textbf{Prompt Template},
        fonttitle=\bfseries,
        boxrule=0.6pt,
        arc=2pt,
        left=3pt, right=3pt, top=3pt, bottom=3pt,
        width=0.98\columnwidth
    ]
    \ttfamily\scriptsize

You are an HPC I/O architecture expert.\\
Your task is to analyze the provided \textcolor{blue}{hybrid JSON context} and map it to the most suitable GekkoFS architecture mode.\\

\#\#\# Knowledge Base\\
\{MODE\_INFO\}\\

\#\#\# Application Context\\
\{APP\_INFO\}\\

\#\#\# \textcolor{blue}{Hybrid Context (Static + Runtime)}\\
\{CONTEXTUAL\_SUMMARY\}\\

\#\#\# \textcolor{orange}{Reasoning Requirements}\\
1. Analyze topology: isolated (N-N) vs shared (N-1).\\
2. Analyze intensity: metadata vs bandwidth.\\
3. Analyze direction: read-dominant vs write-dominant.\\
4. Analyze phase behavior across execution.\\

\#\#\# \textcolor{orange}{Reasoning Strategy}\\
Perform step-by-step reasoning over the provided context and avoid unsupported assumptions.\\

\#\#\# Mode Selection Task\\
Select the layout mode that best matches the workload characteristics.\\

\textbf{Constraint:} Select exactly one from [Mode 1, Mode 2, Mode 3, Mode 4].\\

\#\#\# \textcolor{teal}{Output (JSON Only)}\\
\{
  "selected\_mode": "Mode X",
  "confidence\_score": 0.0-1.0,
  "io\_topology": "N-N or N-1",
  "primary\_reason": "Step-by-step reasoning",
  "risk\_analysis": "Potential trade-offs"
\}

    \end{tcolorbox}
    \caption{Prompt template for hybrid intent inference.}
    \label{lst:prompt_design}
\end{figure}




\section{Evaluation}
\label{sec:evaluation}
\subsection{Experimental Setup}
Experiments are conducted on the Tianhe Exascale Prototype Upgrade System~\cite{tianhe}, utilizing a unified prototype that extends GekkoFS v0.9.2~\cite{gekkofs}  to support the four layout modes defined in Section~\ref{sec:multi_mode_engine}. To characterize the performance trade-offs and optimal use cases of each mode, we evaluate four classes of workloads: (1) \textbf{IOR}\cite{ior2007}, representing large-scale sequential checkpoint/restart I/O dominated by bandwidth and locality; (2) \textbf{FIO}\cite{fio}, capturing small-file random I/O highly sensitive to tail latency and QoS stability; (3) \textbf{MDTest}\cite{mdtest}, testing metadata-intensive operations governed by namespace visibility and coordination logic; and (4) \textbf{Production Kernels} (HACC\cite{haccio}, S3D\cite{S3D}, MADbench2\cite{madbench} ), covering multi-phase I/O and mixed access patterns. This covers traditional HPC simulations, metadata storms, and modern distributed deep learning checkpoints, and the comprehensive workload matrix is shown in Table~\ref{tab:workload_details}. To ensure robust evaluation, we scale each base scenario across multiple configurations, varying node counts (8, 16, 32), I/O block sizes. 
We configure sequential bandwidth benchmarks with 4~MiB transfer sizes, while FIO random-I/O experiments use 4~KiB requests; page caches are flushed prior to read operations to measure true hardware behavior.

\begin{table}[t]
\centering
\caption{Workload configurations for evaluation. We cover an extensive set of workload scenarios across representative HPC applications, spanning diverse I/O patterns (N-N vs. N-1) and intensity levels.}
\label{tab:workload_details}
\resizebox{\columnwidth}{!}{%
\begin{tabular}{@{}lll@{}}
\toprule
\textbf{App} & \textbf{ID} & \textbf{Semantics \& I/O Pattern} \\ \midrule
\multirow{4}{*}{\textbf{IOR}} 
 & Test-A & \textbf{N-N Write}: Indep. file-per-process, Seq. \\
 & Test-B & \textbf{N-1 Read}: Shared file, collision-heavy. \\
 & Test-C & \textbf{Meta-Heavy}: Small segmented R/W. \\
 & Test-D & \textbf{Mixed}: Segmented dynamic R/W access. \\ \midrule
\multirow{4}{*}{\textbf{FIO}} 
 & Test-A & \textbf{N-N Write}: Checkpoint simulation. \\
 & Test-C & \textbf{AI/Meta}: Massive small files, Rand access. \\
 & Test-D & \textbf{Hybrid}: N-1 Write + Rand Read (30\%). \\
 & Test-E* & \textbf{Shared R/W}: Read ratios 10\%, 50\%, 90\%. \\ \midrule
\multirow{3}{*}{\textbf{HACC}} 
 & Test-A & \textbf{N-1 Write}: Large-scale checkpointing. \\
 & Test-B & \textbf{N-1 Read}: Global analysis/restart. \\
 & Test-C & \textbf{Latency}: Small meta ops sensitivity. \\ \midrule
\multirow{3}{*}{\textbf{MAD}} 
 & Test-A & \textbf{N-1 Write}: Collective I/O coordination. \\
 & Test-B & \textbf{N-N Write}: Unique stream throughput. \\
 & Test-C & \textbf{Small I/O}: Mixed data \& metadata. \\ \midrule
\multirow{4}{*}{\textbf{MDTEST}} 
 & Test-A & \textbf{Indep. Meta}: File-per-process (Unique Dir). \\
 & Test-B & \textbf{Shared Meta}: N-1 Dir contention. \\
 & Test-C & \textbf{Deep Tree}: Recursive, namespace stress. \\
 & Test-D & \textbf{2-Phase}: Create then Stat (Cache test). \\ \midrule
\multirow{3}{*}{\textbf{S3D}} 
 & Test-A & \textbf{N-N Write}: Checkpoint burst. \\
 & Test-B & \textbf{Global Read}: Restart pattern. \\
 & Test-C & \textbf{Small I/O}: Latency-sensitive. \\ \bottomrule
\end{tabular}%
}
\end{table}

\subsection{Microbenchmarks: Validating Multi-Mode Layouts}
\label{sec:bbevaluation}

We evaluate \systemname{}'s four layout modes to systematically explore their performance envelopes and validate our central hypothesis: BB performance fundamentally depends on aligning layout assumptions with application I/O semantics. By quantifying performance degradation in mismatched scenarios, we demonstrate multi-mode adaptation is strictly required to surpass monolithic performance ceilings.

\begin{figure}[t]
    \centering
    \includegraphics[width=0.97\columnwidth]{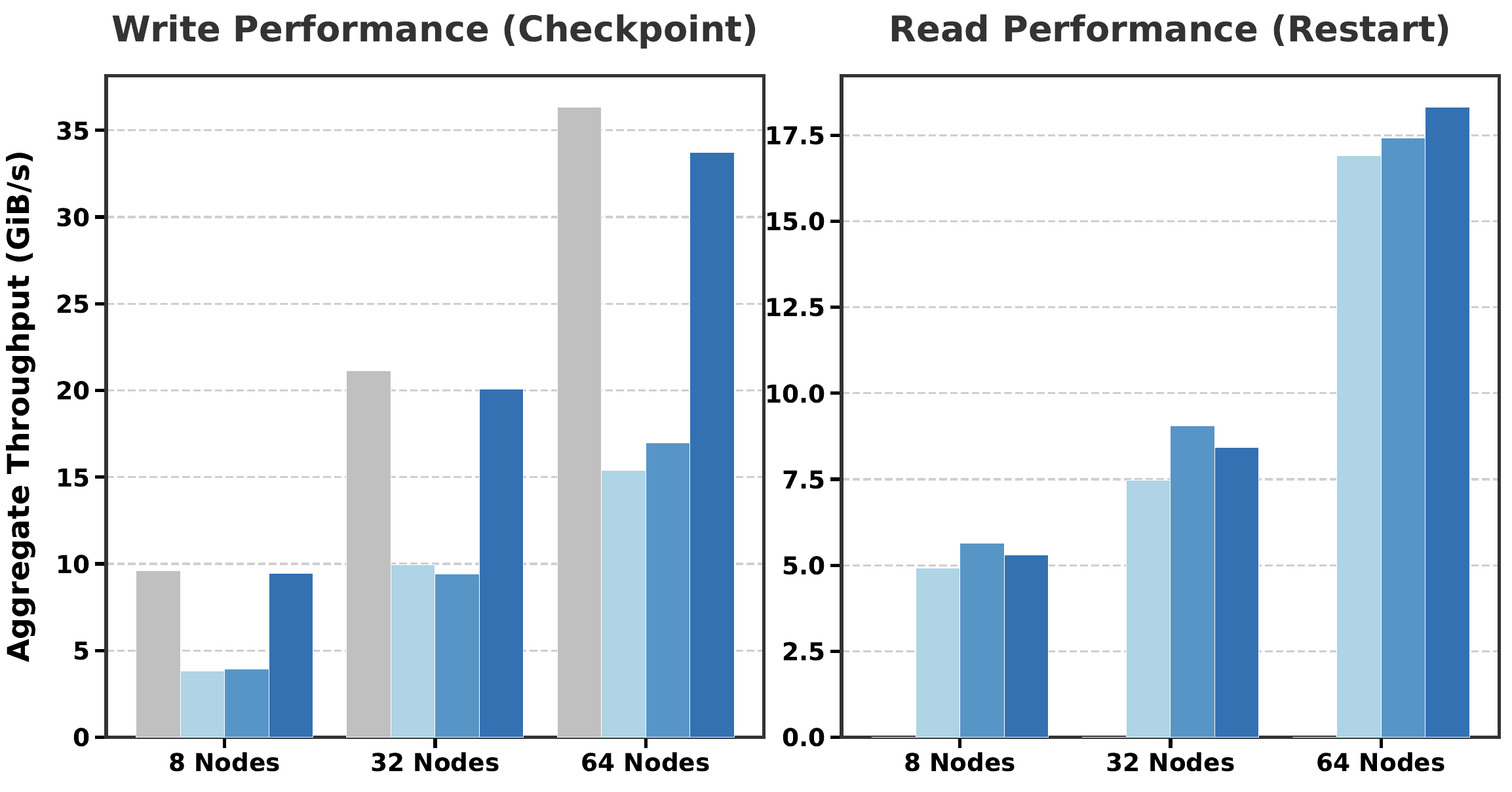}
    \caption{Throughput for checkpoint (write) and restart (read). Mode 1 maximizes write bandwidth via node-local isolation, while Mode 4 balances write performance with the global visibility required for restart operations.}
    \label{fig:ior_bw}
\end{figure}

\paragraph{Bandwidth Analysis in Checkpoint/Restart}
\label{sec:ior_eval}
Sequential I/O highlights the structural tension between write-phase locality and read-phase global visibility. In Fig.~\ref{fig:ior_bw}, Mode~1 and Mode~4 dominate the checkpoint phase. At 64 nodes, Mode~1 achieves 35~GiB/s, demonstrating near-ideal linear scalability by strictly confining I/O traffic to node-local tiers, thus eliminating network coordination for independent $N$--$N$ writes. However, this physical isolation becomes a severe liability during restart, as stranded local data incurs massive cross-node penalties. Conversely, Mode~4 (Hybrid) sustains 17.5~GiB/s by combining local writes with globally visible metadata. This proves monolithic layouts cannot satisfy multi-phase requirements, necessitating composite routing logic like Mode~4.

\begin{figure}[t]
    \centering
    \includegraphics[width=0.98\columnwidth]{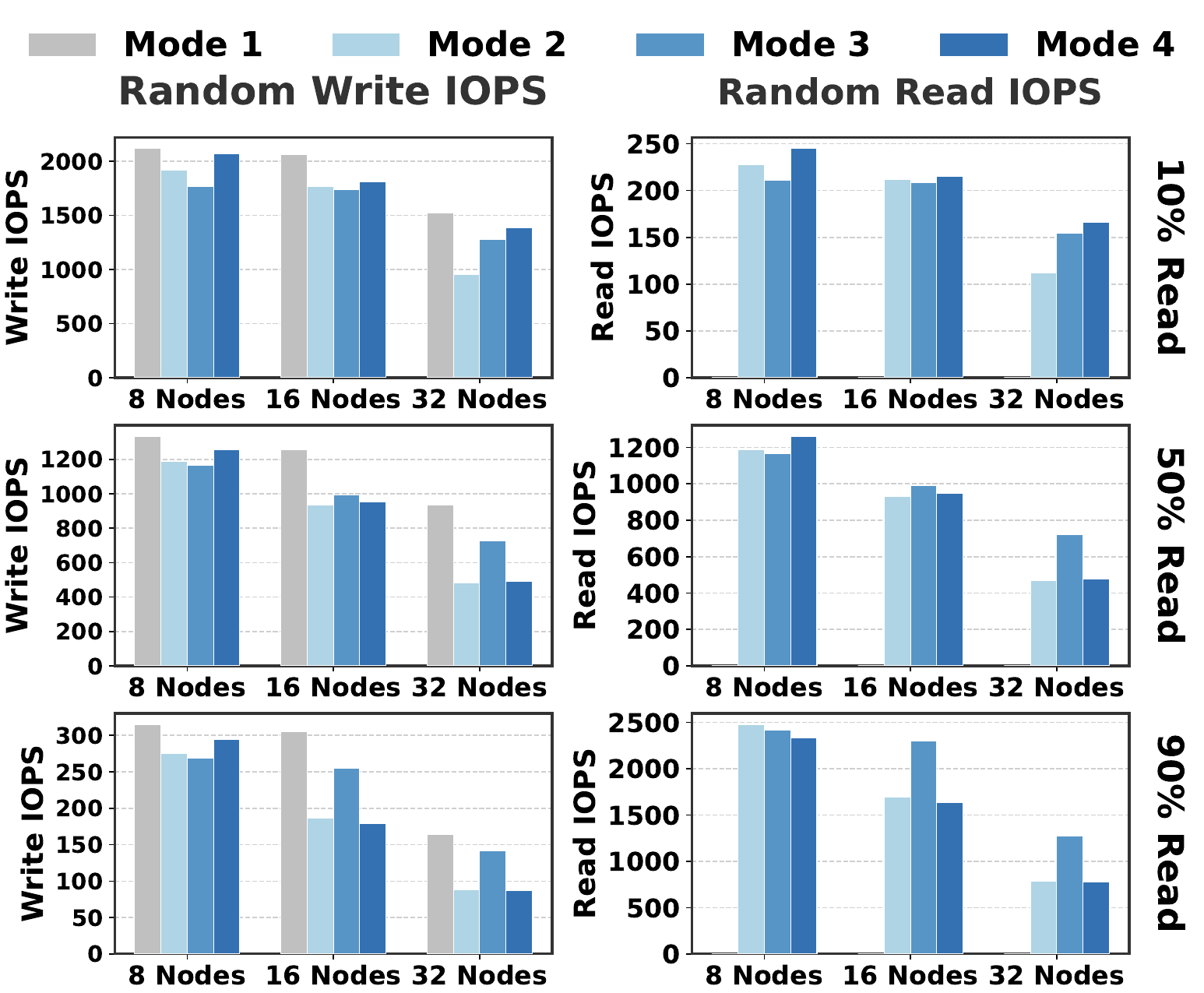}
    \caption{Random I/O scalability measured by IOPS (FIO). Performance comparison across varying read ratios and cluster sizes. Mode~3 provides the most consistent scalability for random access patterns.}
    \label{fig:fio_bw_scaling}
\end{figure}

\begin{figure}[t]
  \centering
  \begin{subfigure}{\columnwidth}
    \centering
    \includegraphics[width=1\linewidth,height=4cm,keepaspectratio]{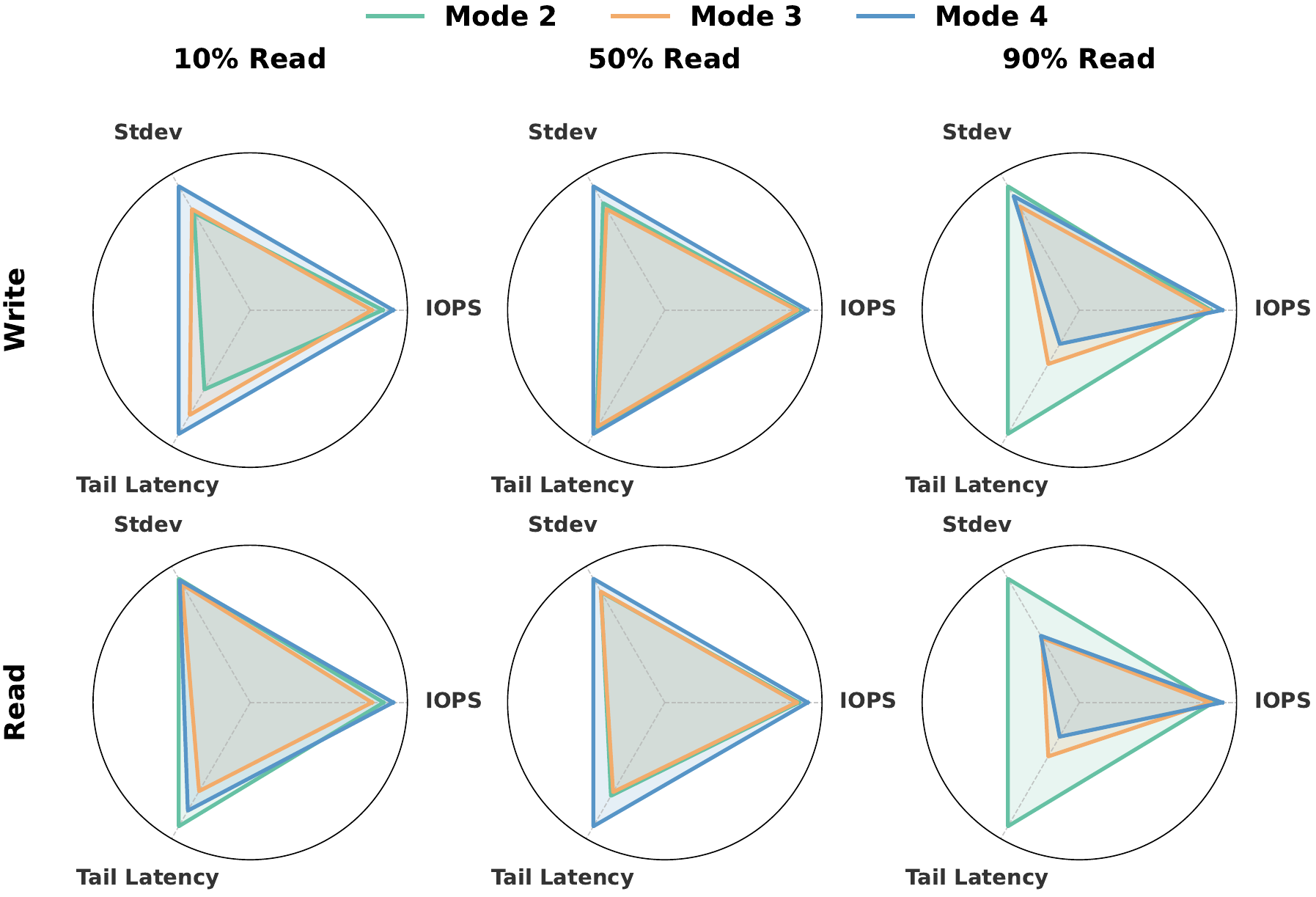}
    \caption{8 Nodes}
    \label{fig:fio_8}
  \end{subfigure}
  \hfill
  \begin{subfigure}{\columnwidth}
    \centering
    \includegraphics[width=1\columnwidth,height=4cm,keepaspectratio]{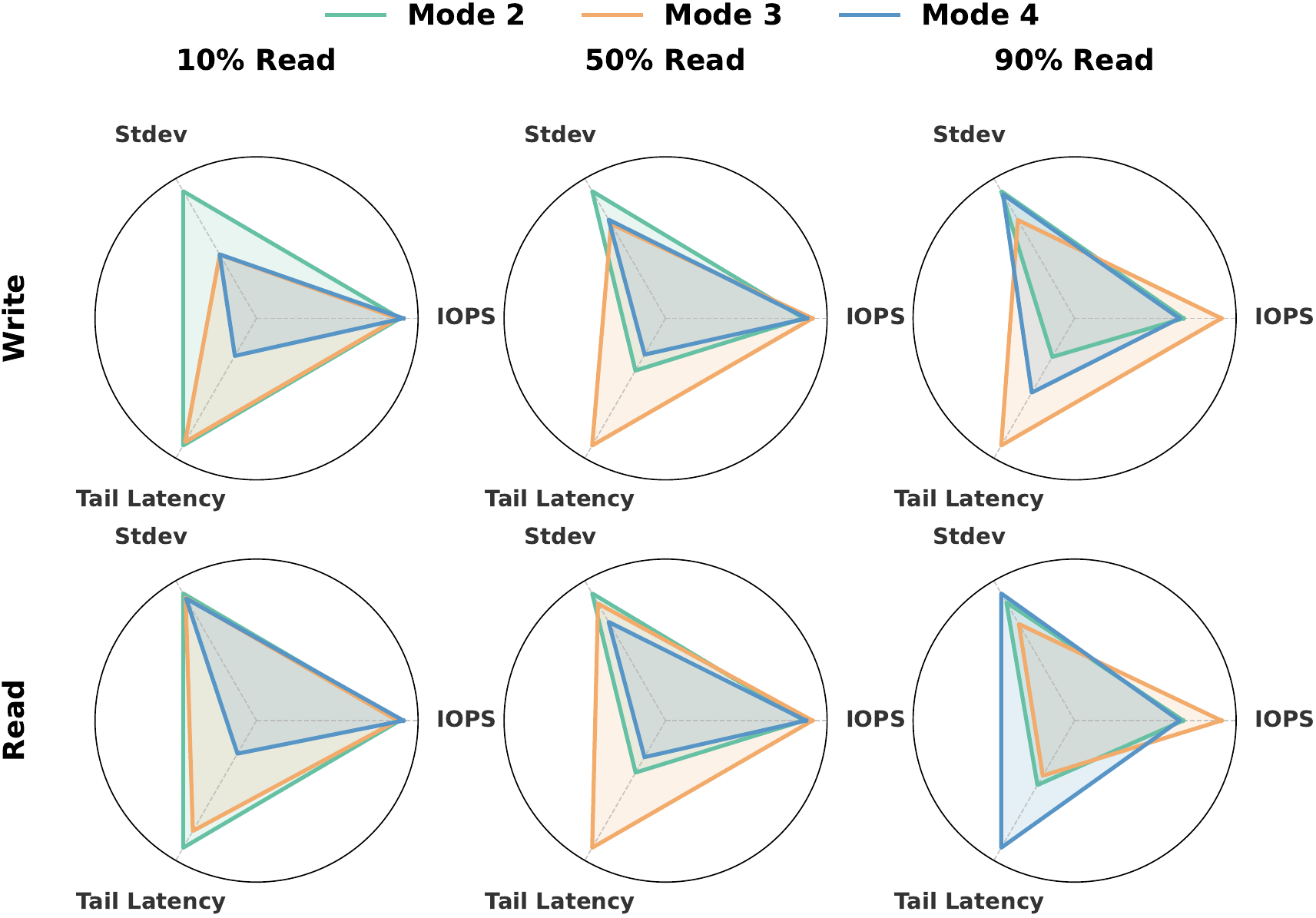}
    \caption{16 Nodes}
    \label{fig:fio_16}
  \end{subfigure}
  \hfill
  \begin{subfigure}{\columnwidth}
    \centering
    \includegraphics[width=1\linewidth,height=4cm,keepaspectratio]{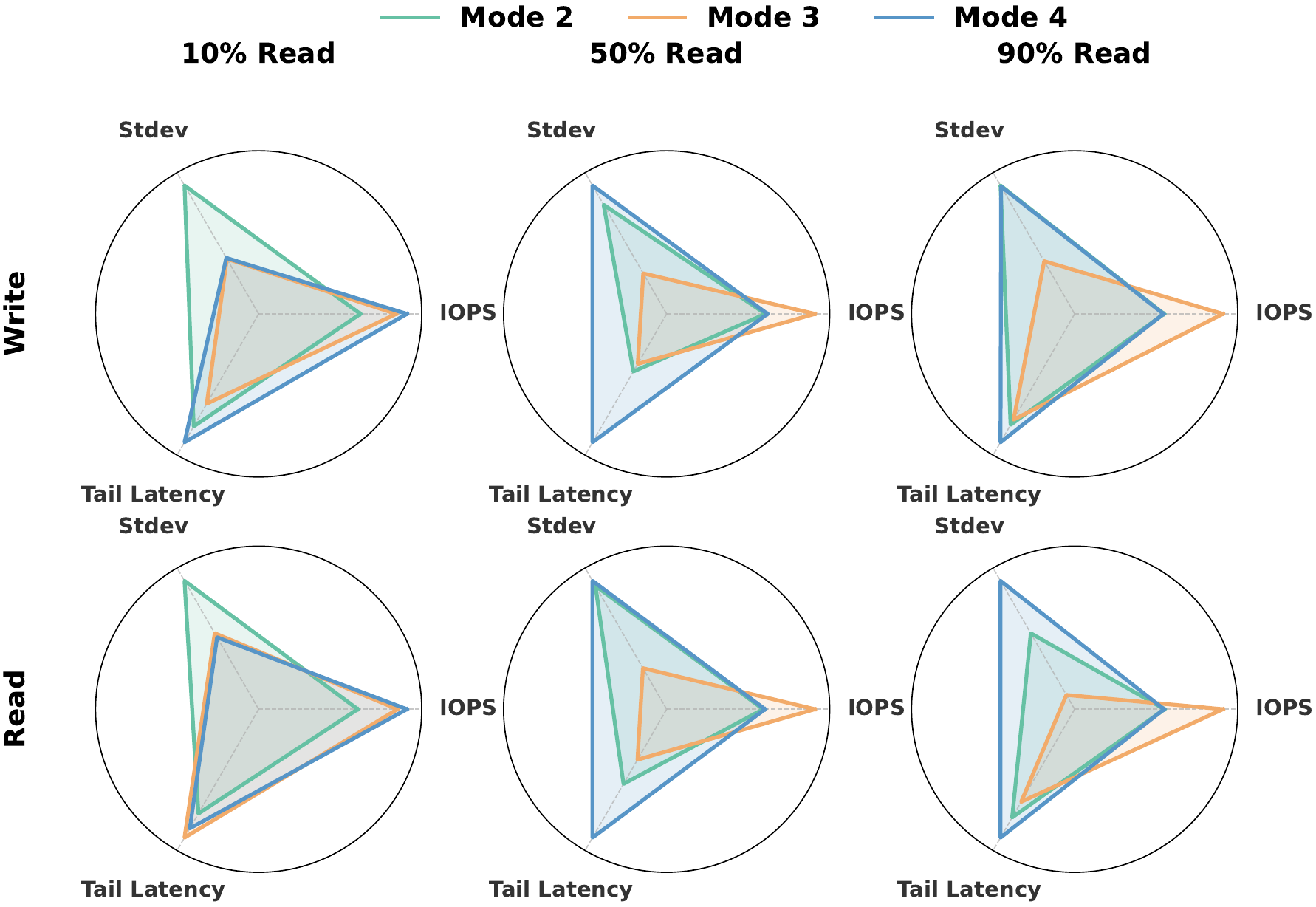}
    \caption{32 Nodes}
    \label{pics/fig:fio_32}
  \end{subfigure}
  \caption{QoS and tail latency analysis. Radar plots show that Mode 2  offers the most stable latency for small I/O, while Mode 3 achieves higher aggregate IOPS at scale.}
  \label{fig:fio_scaling_all}
\end{figure}

\paragraph{Scalability of Random I/O}
\label{sec:fio_eval}
Random I/O analysis (Fig.~\ref{fig:fio_bw_scaling}, \ref{fig:fio_scaling_all}) exposes deep trade-offs among peak IOPS, scalability, and QoS stability. Mode~1's write performance collapses to 164~IOPS under 90\% read ratios at 32 nodes, confirming that excessive locality severely penalizes shared random access. While Mode~4 achieves high IOPS in small clusters, it exhibits severe performance jitter at 32 nodes. Conversely, Mode~2 maintains the lowest standard deviation and the most stable tail latency, demonstrating the value of centralized arbitration for latency-sensitive semantics. Mode~3 delivers the best scaling read performance, reaching 1272~IOPS in high-read scenarios. This divergence confirms that architectural stability and peak random-I/O performance are not always aligned, requiring precise mode matching before execution.

\begin{figure}[t]
    \centering
    \includegraphics[width=0.7\columnwidth]{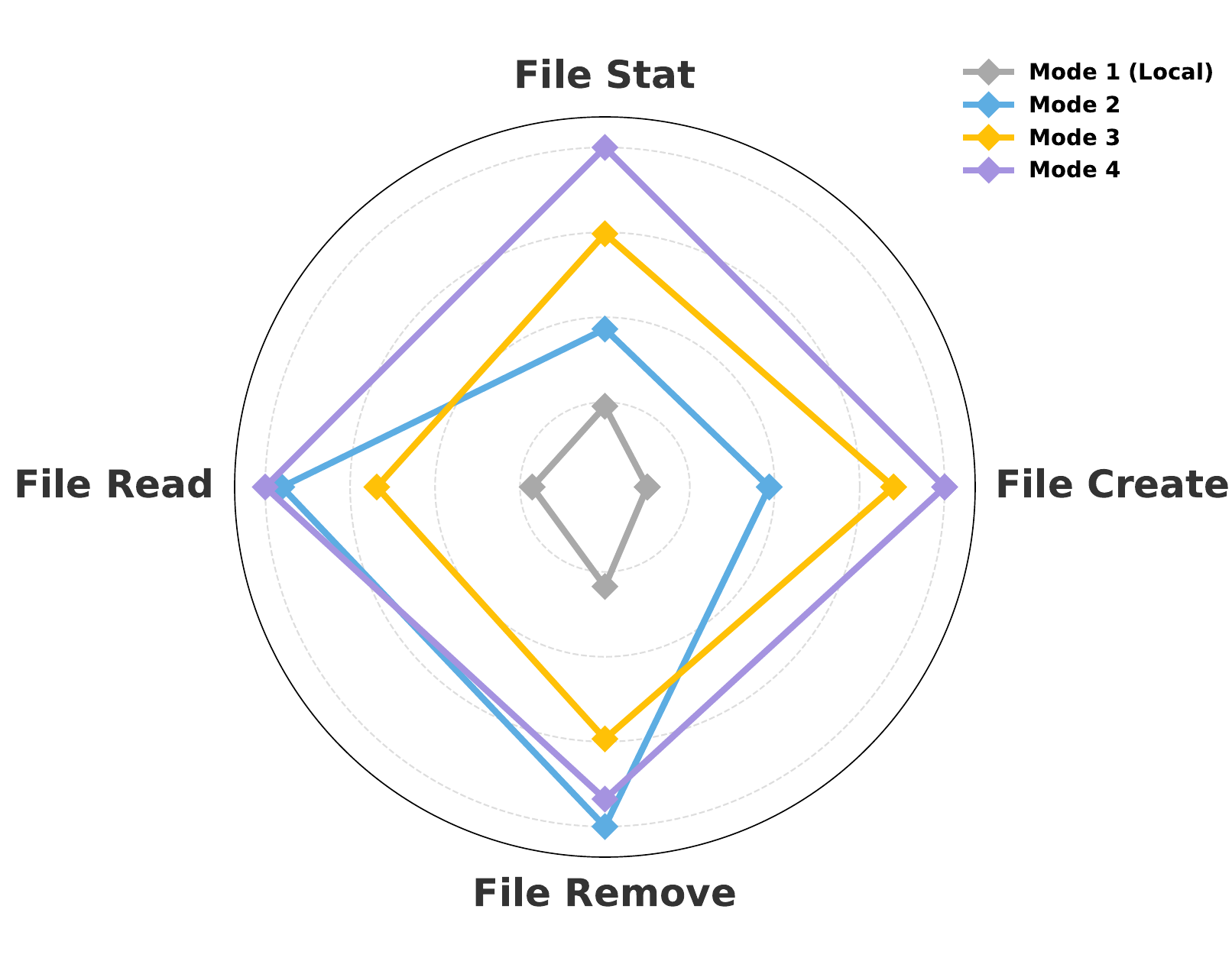}
    \caption{Metadata performance comparison. Mode 2 excels at remove operations due to centralized coordination, while Mode 4 achieves high create throughput via local buffering.}
    \label{fig:mdtest_radar}
\end{figure}

\paragraph{Metadata Operation Performance}
\label{sec:mdtest_eval}
Metadata performance relies heavily on namespace organization. As Fig.~\ref{fig:mdtest_radar} shows, Mode~4 excels in file creation and \texttt{stat} operations via hybrid buffering. However, Mode~2 dominates \texttt{remove} operations and matches Mode~4 in reads, as its centralized metadata subset minimizes distributed locking overhead and accelerates traversals. Conversely, Mode~1 suffers a structural collapse in global tasks, highlighting the catastrophic penalty of physical isolation under shared assumptions. This mismatch confirms layout placement strictly dictates efficiency for namespace-intensive workloads.

\begin{figure}[t]
    \centering
    \includegraphics[width=1\columnwidth]{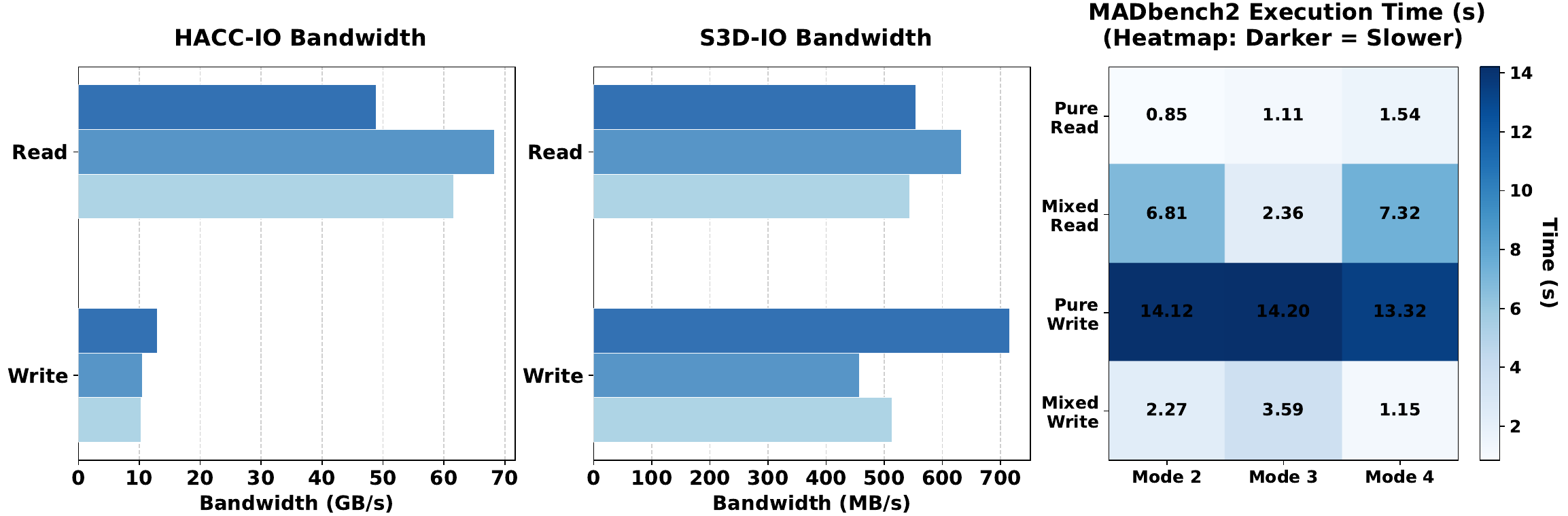}
    \caption{End-to-end performance on HPC kernels. Different kernels prefer different modes, validating the necessity of multi-mode adaptation over a fixed layout.}
    \label{fig:app_validation}
\end{figure}

\paragraph{Validation on Production Kernels and Structural Stability}
\label{sec:app_eval}
Evaluation on representative kernels (Fig.~\ref{fig:app_validation}) confirms no single mode is universally optimal: S3D-IO (write-heavy) aligns with Mode~4; MADbench2 (multi-phase) benefits from Mode~3's decentralized balance; and HACC-IO achieves maximal stability under Mode~2. These end-to-end results confirm that synthetic architectural constraints persist in complex production loads.



\subsection{Accuracy of LLM Intent Inference}
\label{sec:static_eval}

\paragraph{Inference Accuracy Comparison}
We measure decision accuracy against an oracle baseline, defined as the empirically optimal mode determined by exhaustive execution across all layout configurations.

As shown in Table \ref{tab:hybrid_accuracy}, \systemname{} outperforms raditional machine-learning baselines (XGBoost). Despite leveraging historical execution traces and statistical modeling, this ML-based approach still struggles to generalize to complex or unseen multi-phase I/O patterns. In contrast, by leveraging semantic reasoning, \systemname{} significantly improves layout selection accuracy, reaching an optimal accuracy of 91.30\%.

\begin{table}[t]
  \centering
  \caption{Comparison of mode selection accuracy.}
  \label{tab:hybrid_accuracy}
  \begin{tabular}{lc}
    \toprule
    \textbf{Method} & \textbf{Accuracy} \\
    \midrule
    XGBoost & 73.91\% \\
    \systemname{} (Ours) & \textbf{91.30\%} \\
    \midrule
    \multicolumn{2}{l}{\textit{\systemname{} with Different LLMs}} \\
    \midrule
    \quad Qwen3-235B      & \textbf{91.30\%} \\
    \quad Gemini-2.5-Flash   & 86.96\% \\
    \quad DeepSeek-R1     & 73.91\% \\
    \quad GPT-4o          & 73.91\% \\
    \quad Qwen3-32B       & 52.17\% \\
    \bottomrule
  \end{tabular}
\end{table}

\begin{table}[t]
  \centering
  \caption{Ablation study on intent inference components.}
  \label{tab:ablation}
  \begin{tabular}{lc}
    \toprule
    \textbf{Configuration} & \textbf{Accuracy} \\
    \midrule
    \systemname{} (Full Pipeline) & \textbf{91.30\%} \\
    \midrule
    \textit{Context Ablation} & \\
    \quad w/o Runtime (Static Only) & 86.96\% \\
    \midrule
    \textit{Knowledge Ablation} & \\
    \quad w/o App-Ref    & 82.60\% \\
    \quad w/o Mode-Know  & 65.20\% \\
    \bottomrule
  \end{tabular}
\end{table}
Across different LLMs, higher-capacity models consistently achieve better performance. 
These results indicate that layout selection requires non-trivial reasoning over multiple interacting I/O features rather than simple pattern matching. We adopt Qwen3-235B as the default model for analysis.

\paragraph{Ablation Study on Domain Knowledge}

To quantify the contribution of different design choices in \systemname{}, we conduct an ablation study evaluating the impact of the hybrid context and the domain knowledge (Table \ref{tab:ablation}).

First, for context extraction, removing lightweight runtime signals (w/o Runtime) and relying solely on static inference drops the accuracy to 86.96\%. This confirms that while static artifacts provide a robust baseline, runtime execution evidence is essential to resolve complex semantic ambiguities.
Second, for domain knowledge, removing mode architectural descriptions (w/o Mode-Know) leads to the most significant degradation (dropping to 65.20\%), indicating that understanding the structural trade-offs of the storage backend is the critical factor for correct layout mapping. Furthermore, removing application-level semantic guidelines (w/o App-Ref) reduces accuracy to 82.60\%, highlighting the importance of cross-layer reasoning in bridging application behaviors with storage systems.

\paragraph{Cost Analysis}In \systemname{}, the end-to-end decision pipeline consists of static extraction, a single execution probe, LLM inference, and mode activation. Specifically, when evaluated using Qwen3-235B, the LLM inference phase accounts for the majority of the system-intrinsic overhead, taking 33.0\,s (with a 95th percentile of 51.3\,s), and each decision request consumes $\sim$9.4k input tokens and generates $\sim$1.1k output tokens.

Table \ref{tab:cost_analysis} compares this overhead against traditional black-box ML tuning paradigms. ML methods suffer from severe cold-start problems, typically requiring $10^2$--$10^3$ offline training executions and 10--100 full profiling runs prior to deployment. In contrast, by relying on execution probes and a single inference call, \systemname{} avoids exhaustive parameter sweeps and the hundreds of repeated full-length runs required by traditional ML. Consequently, this optimization tax is negligible over typical HPC jobs, preserving high deployment practicality in production environments.

\begin{table}[t]
\caption{Cost comparison across optimization paradigms.}
\label{tab:cost_analysis}
\centering
\begin{tabular}{@{}lcc@{}}
\toprule
\textbf{Metric} & \textbf{Black-box ML} & \textbf{\systemname{} (Ours)} \\ \midrule
Offline training cost & $10^2$--$10^3$ runs & \textbf{0} \\
Pre-execution profiling & 10--100 full runs & \textbf{1--2 probes} \\
Decision latency & $\sim$ms & $\sim$48.5 s \\
Feature source & Runtime statistics & Hybrid (Static + Probes) \\
Search space & Parameter tuning & Structural layout \\
\bottomrule
\end{tabular}
\end{table}

\subsection{End-to-End Performance Evaluation}

\paragraph{Overall Performance Speedup}
\begin{figure}[t]
    \centering
    \includegraphics[width=0.9\columnwidth]{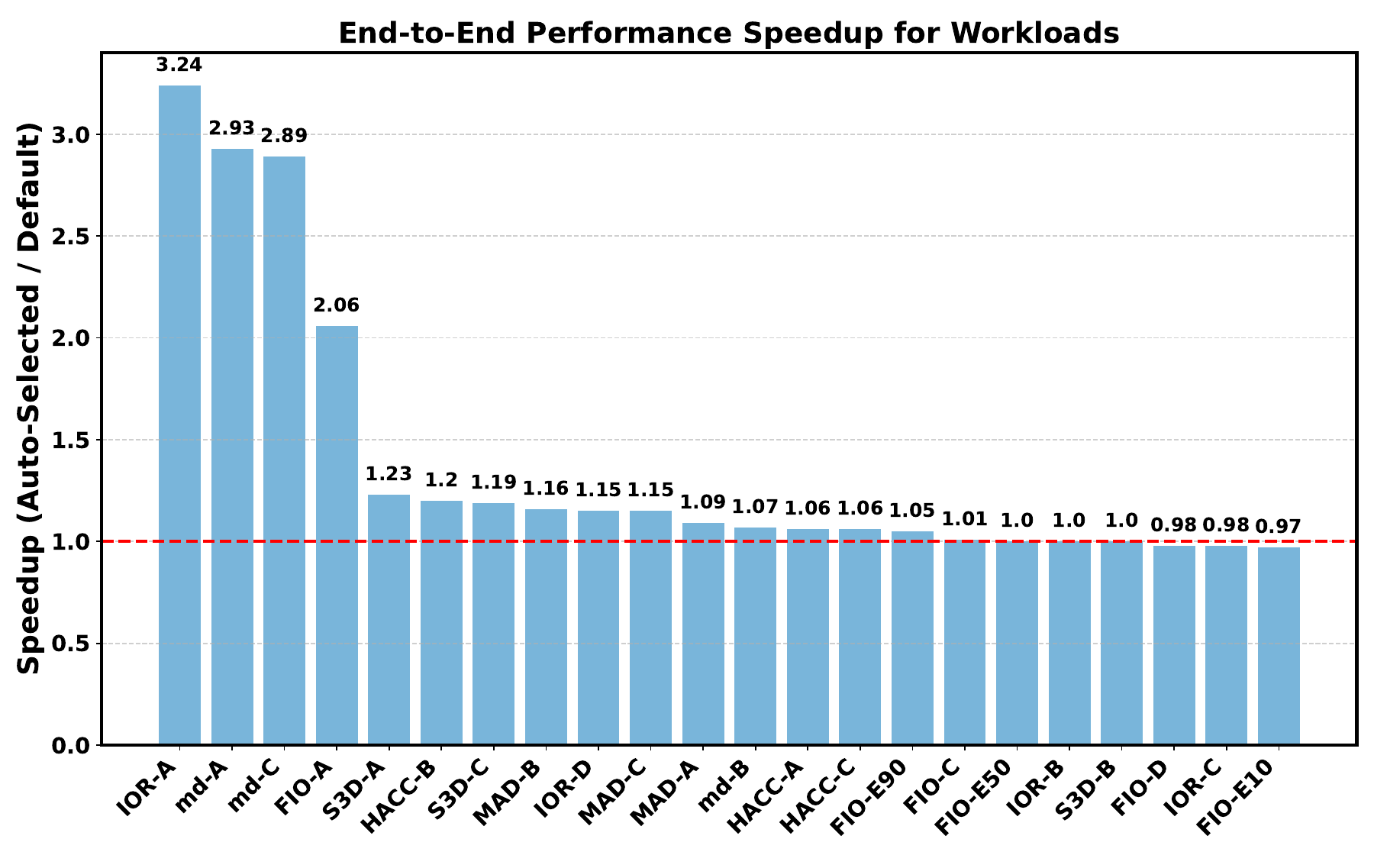}
    \caption{Performance speedup of Proteus. Comparison against the default GekkoFS baseline across an extensive set of diverse workloads, showing up to 3.24$\times$ improvement.}
    \label{fig:final_bar}
\end{figure}
Each benchmark contains multiple subtests (e.g., IOR-A/B/C/D), as listed in Table~\ref{tab:workload_details}. Together, these extensive scenarios comprehensively cover file-per-process ($N$--$N$), shared-file ($N$--$1$), metadata-intensive, and mixed phases, spanning write-dominated, read-dominated, and hybrid HPC I/O behaviors.
Figure~\ref{fig:final_bar} shows that \systemname{} achieves clear gains on bottlenecked cases and remains near baseline on already well-aligned cases. It reaches 3.24$\times$ (IOR-A), 2.93$\times$ (mdtest-A), and 2.89$\times$ (mdtest-C), while maintaining robust gains of 1.15$\times$--1.23$\times$ on shared-access kernels such as S3D and HACC-B.



\paragraph{Comparison with Parameter Tuning and State-of-the-Art Burst Buffers}
We compare \systemname{} against three representative baselines: (1) OPRAEL~\cite{liu2023optimizing}, a state-of-the-art auto-tuning framework that utilizes an ensemble of machine learning models (e.g., Random Forest, XGBoost) to search for optimal storage configurations over GekkoFS; (2) UnifyFS, a node-local write-optimized burst buffer; and (3) CodepFS, a pattern-aware distributed burst buffer. For a fair comparison, UnifyFS and CodepFS were deployed using their officially recommended configurations and best practices. 

As shown in Fig.~\ref{fig:final_bar}, OPRAEL represents the performance limit of traditional ML auto-tuning over a fixed architecture. It generally falls short of \systemname{}'s layout-level optimizations because parameter tuning alone cannot overcome foundational architectural mismatches. For example, in metadata-intensive workloads (e.g., md-A and md-C), OPRAEL fails to resolve inherent distributed locking bottlenecks, yielding limited gains. Similarly, specialized burst buffers excel only in their target domains: UnifyFS achieves excellent performance on $N$--$N$ write-heavy workloads (e.g., 3.05$\times$ on IOR-A), perfectly matching its architectural design goals. However, it severely degrades on shared-access and metadata-heavy cases (e.g., MAD-B). We emphasize that this degradation is fundamentally architectural (e.g., lack of global metadata arbitration), perfectly aligning with our core argument that fixed layouts impose rigid performance ceilings. 
In contrast, by dynamically reshaping its underlying layout prior to execution, \systemname{} successfully bypasses these rigid limits. It consistently achieves near-optimal upper bounds in favorable scenarios (e.g., 3.24$\times$ on IOR-A) and delivers massive speedups in typically bottlenecked cases (e.g., 2.49$\times$ on md-A and 2.07$\times$ on md-C).

\begin{figure}[t]
    \centering
    \includegraphics[width=1\columnwidth]{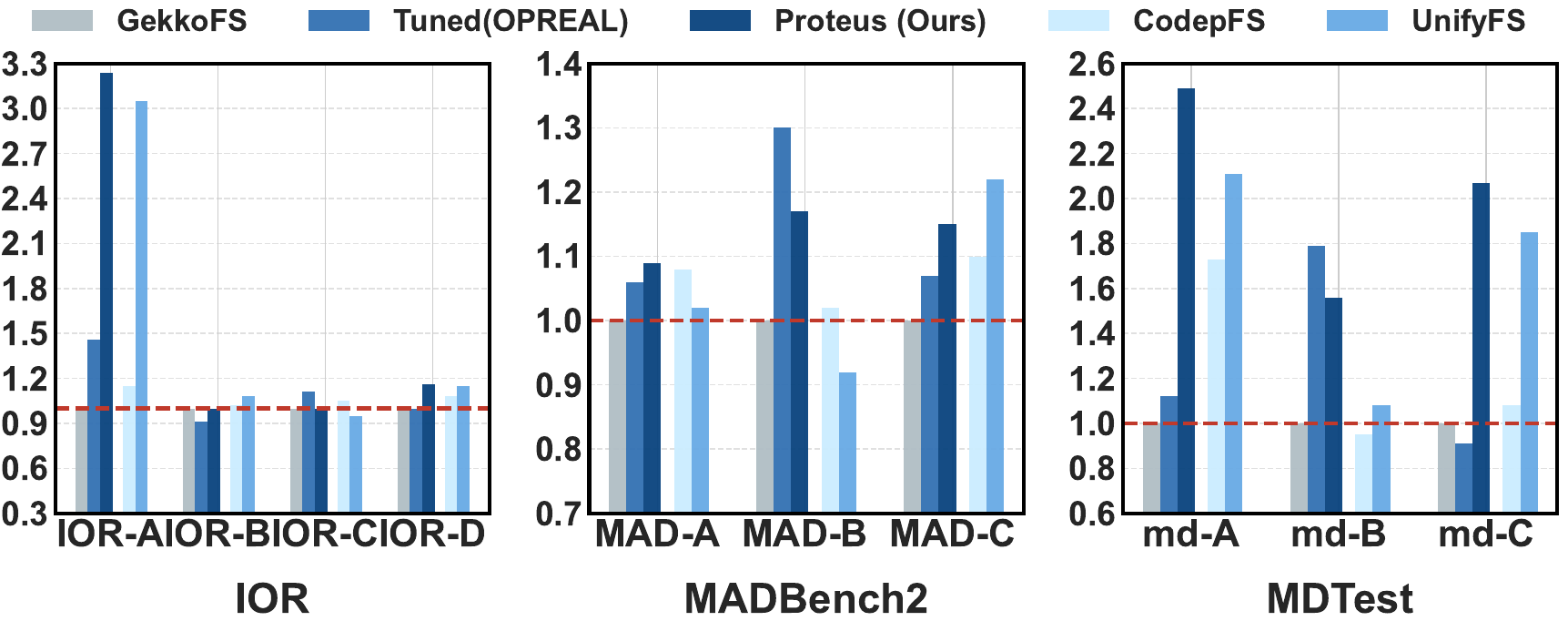}
     \caption{End-to-end performance comparison across different systems.}
    \label{fig:final_bar}
\end{figure}

\begin{figure}[t]
    \centering
    \includegraphics[width=0.98\columnwidth]{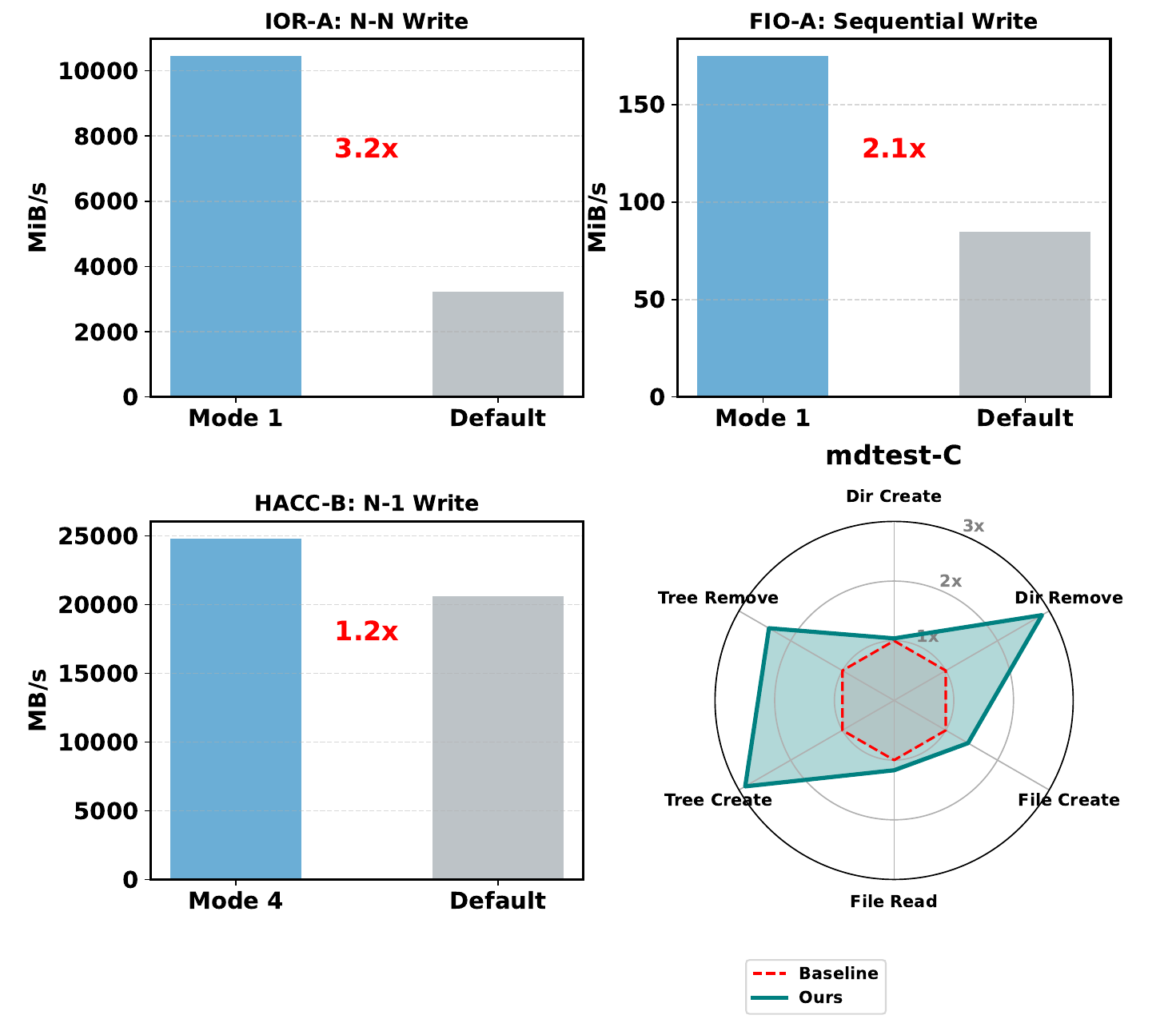}
    \caption{Detailed performance of case study analysis.}
    \label{fig:final_4}
\end{figure}

\paragraph{Representative Case Studies: From Reasoning to Performance}
We select three scenarios to verify the causal link between LLM reasoning and final speedup. Figure~\ref{fig:final_4} reports detailed results.

(1) Exploiting isolation for hardware-native bandwidth (IOR-A/FIO-A):
The LLM parses \texttt{-F} and \texttt{-a POSIX}, identifies an isolated $N$--$N$ pattern, and selects Mode~1. The decision removes unnecessary consistency traffic and drives throughput to 10,457~MiB/s.

(2) Aligning write bursts with global consistency (HACC-B):
The model detects $N$--$1$ shared access and write-dominant checkpoint bursts, then selects Mode~4 (Hybrid) to combine local write bandwidth with global namespace visibility, reaching 24,807~MB/s write throughput.

(3) Mitigating metadata storms via centralization (mdtest):
The model detects shared-directory contention (e.g., \texttt{-d} without \texttt{-u}) and selects Mode~2 to centralize metadata arbitration, reducing lock contention under concurrent namespace operations.


\section{Conclusion and Future Work}
\label{sec:conclusion}

This paper presents \systemname{}, a multi-mode burst buffer system that eliminates structural performance ceilings by aligning storage layouts with application I/O intent. Rather than relying on data-heavy ML auto-tuning that requires exhaustive profiling, \systemname{} introduces a training-free, LLM-guided reasoning paradigm. By synergizing static application artifacts with lightweight runtime probes, it accurately infers I/O semantics to dynamically instantiate the optimal layout at job startup. Evaluation shows \systemname{} achieves 91.30\% layout selection accuracy and delivers up to 3.24$\times$ end-to-end speedups, consistently matching or exceeding specialized monolithic burst buffers.

Future work focuses on following directions: (1) fine-tuning compact, locally-hosted LLMs (e.g., 7B–32B parameters) using Proteus's curated hybrid contexts to satisfy HPC privacy and reduce inference latency without sacrificing accuracy; (2) extending Proteus to support sub-job, phase-level layout transitions for highly dynamic multi-physics simulations; and (3) implementing an automated feedback loop to continuously refine the LLM’s knowledge base via historical traces, enabling lifelong learning for adaptive HPC storage.

\bibliographystyle{IEEEtran}
\bibliography{ref}

\end{document}